# Predictions for solar flares activity in solar cycle 25

Eleni Petrakou[*]

*Athens, Greece*

*29 July 2020*

***Abstract:*** *The prediction of the evolution of individual solar cycles is a developing field, faced with divergence of forecasts even for a few years in the future. Specifically for solar flares, long-term modeling is practically absent even in rough terms. In this article a forecast for the evolution of solar cycle 25 in terms of solar flares activity is presented. It is derived from an existing phenomenological model based on the coupling of an internal solar component and a planetary component. In addition to competent temporal resolution, the predictions are characterized by features which both differentiate the model from other space climate forecasts and make it falsifiable.*

**1. Introduction**

Forecasting the next sunspot cycle has been a central activity of space weather and space climate research, with a large number of different approaches taken over the last decades; however, the forecasts from different prediction models still diverge significantly[1]. Moreover, forecasts for the solar cycle in terms of solar flares, instead of sunspots, are practically absent, although flares are both more solid and impactful events.

In this article a forecast for the solar cycle 25 is presented in terms of solar flares activity, as derived from the model for long-term solar evolution in [2], which reconstructed the last three cycles to a promising degree using a minimal set of assumptions.

In [2] the model was extended temporally over the next years, thus providing a forecast for the evolution of solar cycle 25. However, the actual start of the new cycle in late May 2020 calls for an updated prediction, which is presented here. A favorable comparison with the previous prediction for the start of the cycle can also be made.

Section 2 gives a brief overview of the used model and discusses the start of the new cycle and how it led to the update. Section 3 presents the predictions for the new cycle in more detail. Section 4 concludes with a brief summary.

---

[*] email: eleni@petrakou.net

## 2. Start of the new cycle

In the following, "flares" will refer to solar flares of X-ray flux intensity classes M and X. Flares belonging to the solar cycle 25 ("C25") are defined as coming from active regions belonging to C25, according to their standard definition. The start of a cycle is defined as the date with its first flare. The analysis was performed with the ROOT toolkit[3]; the flare records comprise the measurements of the NOAA SMS and GOES satellites[4]; planetary positions and sunspots records were obtained from NASA's HelioWeb[5] and the Royal Observatory of Belgium[6] respectively.

The used model consists of the overlay of two kinds of Gaussian distributions, which are repeatedly centered on specific dates. These Gaussians were derived from the data of solar cycle 21 and are the same for every cycle. One of the two distributions represents the relative ecliptic motion of the planets Jupiter and Saturn, and it is repeatedly centered on the dates of their alignments. The other represents an internal solar component, and it is repeatedly centered on the dates of the temporal middle of each cycle. Full details can be found in [2].

The model can be extended over the coming years by placing the centers of the Gaussian distributions on the appropriate dates. In [2], published in 2018, this had to involve the estimation of the temporal middles of cycles 24 and 25. This estimation is based on the average difference between the solar cycle duration and the two planets' half synodic period, which results in the progressive increase of the temporal distance between the two distributions. Thus, by using the average of this increase, the temporal middle of cycle 24 was previously estimated to fall on 2015/05/28.

Under this model, the end of a cycle is defined by the start of the next one, i.e. the date of its first flare. The first M-class flare from a region belonging to C25 appeared on 2020/05/29. This places the actual temporal middle of cycle 24 on 2015/03/25, only 64 days earlier than the estimation. Likewise, in a previous preprint version of this article the prediction for the most likely time range for the first C25 flare was "in the weeks following [late March 2020]".

Knowing the actual date for the middle of cycle 24, the forecast was updated by placing appropriately the corresponding Gaussian; it was also updated in the placement of the one for C25, and by calculating an improved uncertainty associated with this estimation. Overall, the change in the forecast is small, due to the time span of cycle 24 being close to the previous prediction. The resulting expansion over the next years is shown in Figure 1, along with the model and the observations for the previous cycles. The following Section elaborates on the predictions.

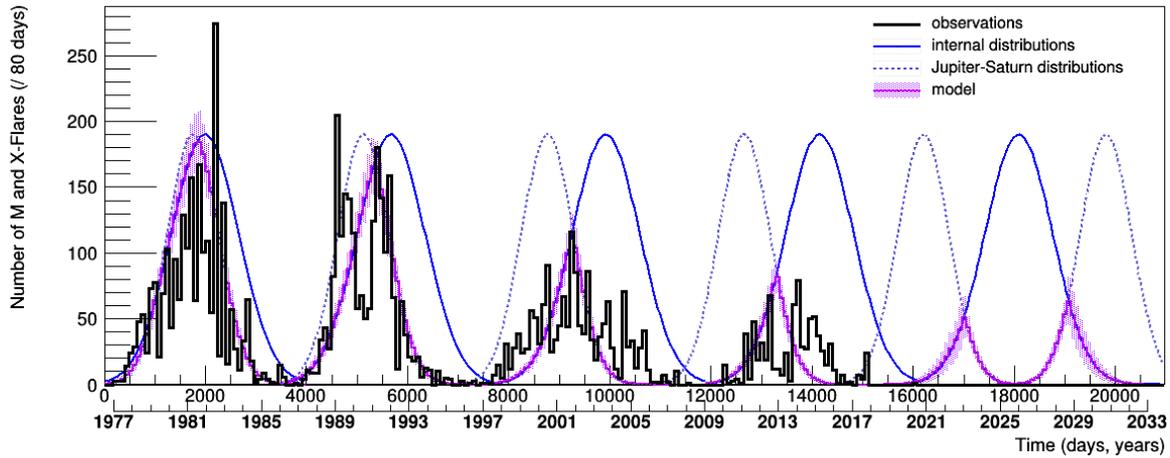

Figure 1. Observations and model for solar flares activity in cycles 21-25 as a function of time: observed number of flares (black) overlaid with the model distributions (purple). The two types of Gaussian components centered on their respective dates are also shown. (The time range starts on 1976/06/30.)

## 3. Forecast

Figure 2 is a zoom-in on the projected time range of C25. The internal Gaussian component is plotted as solid curve and the planetary component as dashed curve; their common areas provide the prediction of the counts of flares as a function of time. The features of the prediction are discussed below along with further explanations.

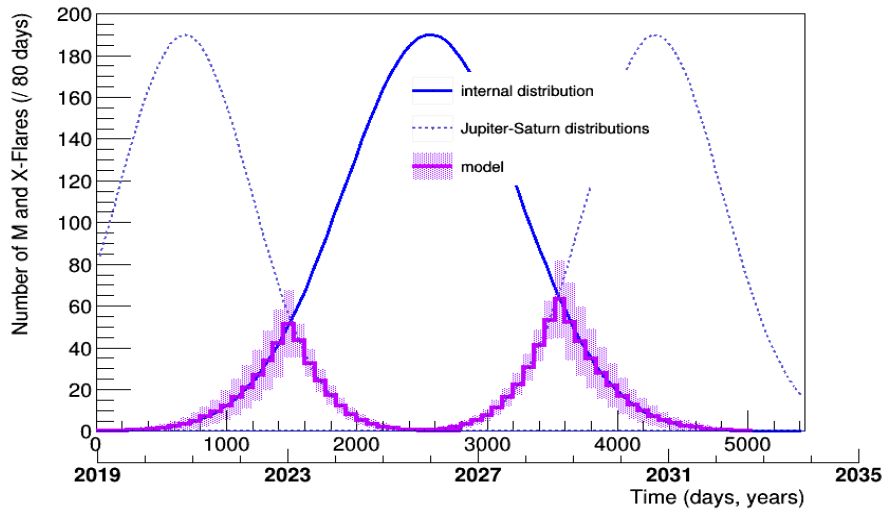

Figure 2. Predicted evolution of solar flares activity in C25 as a function of time (purple histogram) with associated uncertainties. The curves represent the two components discussed in the text (solid: internal; dashed: planetary component). The upper x-axis shows the number of days since the beginning of year 2019.

**Evolution:** The general activity will follow the plotted distribution, i.e. it will be rather low, spread-out, and marked by two distinct ranges. More specifically, it will increase until roughly the beginning of year 2023 and subsequently decrease until a cease around the end of 2025; a distinct second period of activity will follow, along roughly the second part of the plotted histogram.

The two separate stages of activity are due to the presence of two planetary components in the time range covered by the internal solar component: as mentioned above, the planetary distributions are centered on the dates of the alignments of the two planets, which leads to this feature during the upcoming cycle. (In the four fully recorded cycles this has occurred only for a short time towards the end of 24; during that time weaker flares activity indeed showed up anew, after it had ceased completely.)

**Amplitude:** The maximum amplitude is expected to be around 50-60 flares per 80 days, corresponding to the two peaks in the histogram.

**Duration:** With the currently available data, and by estimating the middle of cycle 26 as described above, the start of cycle 26 is placed around the middle of 2031 (not shown in the plots). This makes C25 about 11 years long.

**Overall intensity:** Comparable to that of cycle 24 (see also Figure 1).

**Comparison to historical sunspot minima:** In terms of sunspots, C25 will be similar to the two historical "local minima" in cycles 6 and 14. (This is the only prediction relevant to sunspots.)

The reason is related to the presence of two planetary components. In general, the model cannot be extended to times before cycle 21, since accurate records of flares are needed for the calculation of the temporal middles of cycles in terms of flares. However, a coarse extension can use the middles of the sunspot cycles instead. In that case this "double overlap" appears only in cycles 6 (Dalton minimum) and 14 (Gleissberg minimum); this is compatible with the prediction of a weak and spread-out cycle.

Finally, the uncertainty accompanying the histogram is the quadratic sum of statistical and systematic uncertainties. The systematic uncertainty has two sources: the estimation of the temporal middle of C25, as described above, and the data binning choices. Details of their derivation are found in [2]; the uncertainty due to the temporal middle estimation was updated with the knowledge for cycle 24.

## 4. Conclusions

This article presents a forecast for the temporal evolution of solar cycle 25 and its characteristics in terms of solar flares activity. The forecast is derived from a phenomenological yet deterministic model, based on a synergy between internal solar activity and the relative motion of the two largest planets.

The unique features of the predictions, most notably a clear division into two stages of activity, differentiate the underlying model from other space climate prediction methods, in addition to making it falsifiable. The recent appearance of flares belonging to the new cycle was compatible with the existing prediction about its onset, and led to the presented update.

Finally, it can be noted that planetary influence might seem like an unlikely factor of solar modulation, but it is one of the few conceivable permanent factors which could perturb solar activity either spatially or temporally.